\documentclass[useAMS,usenatbib]{mn2e}
\usepackage{times}
\usepackage{graphicx}
\usepackage{amssymb}
\usepackage[mathscr]{eucal}

\title{Cluster mass estimation through Fair galaxies}
\author[Guido Cupani, Marino Mezzetti, Fabio Mardirossian]
{Guido Cupani$^1$\thanks{E-mail: cupani@oats.inaf.it.}
Marino Mezzetti$^{1,2}$
and Fabio Mardirossian$^{1,2}$\\
$^1$ Dipartimento di Astronomia, Universit\`a degli studi di Trieste, via Tiepolo 11, I-34143 Trieste, Italy\\
$^2$ INAF - Istituto Nazionale di Astrofisica, via Tiepolo 11, I-34143 Trieste, Italy
}
\pagerange{\pageref{firstpage}--\pageref{lastpage}}
\pubyear{2010}
\date{Accepted 2009 December 2. Received 2009 November 30; in original form 2009 October 15}

\newcommand{\bi}{\begin{itemize}}
\newcommand{\be}{\begin{equation}}
\newcommand{\ei}{\end{itemize}}
\newcommand{\ee}{\end{equation}}

\newcommand{\f}{\frac}
\newcommand{\p}{\item}

\begin{document}

\maketitle

\begin{abstract}
We analyse a catalogue of simulated clusters within the theoretical framework of the Spherical Collapse Model (SCM), and demonstrate that the relation between the infall velocity of member galaxies and the cluster matter overdensity can be used to estimate the mass profile of clusters, even though we do not know the full dynamics of all the member galaxies. In fact, we are able to identify a limited subset of member galaxies, the `fair galaxies', which are suitable for this purpose. The fair galaxies are identified within a particular region of the galaxy distribution in the redshift (line-of-sight velocity versus sky-plane distance from the cluster centre). This `fair region' is unambiguously defined through statistical and geometrical assumptions based on the SCM. These results are used to develop a new technique for estimating the mass profiles of observed clusters and subsequently their masses. We tested our technique on a sample of simulated clusters; the mass profiles estimates are proved to be efficient from $1$ up to $7$ virialization radii, within a typical uncertainty factor of 1.5, for more than 90 per cent of the clusters considered. Moreover, as an example, we used our technique to estimate the mass profiles and the masses of some observed clusters of the Cluster Infall Regions in the Sloan Digital Sky Survey catalogue. The technique is shown to be reliable also when it is applied to sparse populated clusters. These characteristics make our technique suitable to be used in clusters of large observational catalogues. 
\end{abstract}


{\setlength\arraycolsep{1pt}

\section{Introduction}\label{sec:intro} 

The clusters of galaxies are the largest dynamically relaxed structure we observe in the Universe. The knowledge of their dynamics and mass distribution is strategic in modern cosmology, since it constrains different cosmological models \citep{WF,BC,Borgal97}. Several methods are currently used to estimate the cluster masses and mass profiles, based on the dynamical analysis of the member galaxies \citep{K,RG,DG,Gal}, on the observations of the X-ray emitting gas \citep{CHM,Borgal}, and on the gravitational lensing effects \citep{GN,NB}. However, the estimation of cluster masses remains difficult, mainly because most of the matter in clusters is actually dark matter, which cannot be easily observed.

Within this framework, the dynamics of the cluster outskirts is particularly relevant. In the oustkirts, the matter is not yet in equilibrium and is affected by an overall infall motion towards the cluster centre. As discussed by \citet[hereafter CMM08]{CMM08}, the dynamics of this ``non-equilibrium'' region turns out to be quite well described by the Spherical Collapse Model (SCM; \citealt{GG,S,P76,G,P80}). CMM08 focused on the turnaround radius $r_\rmn t$ (i.e. the radius where the mean velocity of the infalling matter balances the Hubble flow) and studied the dependence of the overdensity $\delta$ on the radial coordinate $r$ in the region surrounding $r_\rmn t$. They demonstrated that the large majority of clusters are compatible with a single mass profiles when their mass $M$ is expressed in units of the turnaround mass $M_\rmn t$.

The results of CMM08 provide a description of the overall shape of the cluster mass profile in the non-equilibrium region. In principle, a way to reconstruct the individual features of these profiles is provided by the relation between the overdensity and the velocity of the infall motion in clusters (\citealt{RG}):
\be\label{eq:vr_delta}
\omega_\rmn{inf}\equiv\f{v_\rmn{inf}}{H_0 r}\simeq\Omega_{\rmn M,0}^{0.6}F(\delta).
\ee
Here $v_\rmn{inf}$ is the infall velocity (i.e. the bulk peculiar velocity of matter towards the cluster centre, defined as positive when directed inwards), $H_0$ is the present-day Hubble parameter, $\Omega_{\rmn M,0}$ is the present-day matter density parameter and $F$ is an analytical function of the overdensity $\delta$ (CMM08; \citealt{Y,VD}). If $v_\rmn{inf}$ and $F$ are known, one can compute the mass profile by inverting equation (\ref{eq:vr_delta}):
\begin{eqnarray}\label{eq:M_inf}
M(r)&=&\f{4}{3}\pi\rho_{\rmn{cr},0}\Omega_{\rmn M,0}r^3\left[1+\delta(r)\right]\nonumber\\
&=&\f{4}{3}\pi\rho_{\rmn{cr},0}\Omega_{\rmn M,0}r^3\left\{1+F^{-1}\left[\Omega_{\rmn M}^{-0.6}\omega_\rmn{inf}(r)\right]\right\},
\end{eqnarray}
where $\rho_\rmn{cr}$ is the critical matter density, and $F^{-1}$ is the inverse function of $F$. 

To compute the mass profile via equation (\ref{eq:M_inf}) one must reconstruct the overall infall pattern of clusters using the member galaxies as tracers. Such reconstruction may result to be difficult for two reasons:
\begin{enumerate}
\p the radial velocities of member galaxies are affected by the presence of local substructures inside the clusters; therefore, the values of $v_\rmn{inf}$ are distributed along $r$ in a blurred band (see also CMM08 for details) and
\p the infall velocities of galaxies can not be easily inferred from observations, owing to projection effects.
\end{enumerate}
Nevertheless, in the present work we wish to show that the infall velocity approach can be used to estimate the overdensity and mass profiles of galaxy clusters. We focus on the dynamical quantities of galaxies which can be directly inferred from observation, namely the sky-plane distance from the cluster centre, $r_\rmn{sp}$, and the line-of-sight velocity, $u_\rmn{los}$ (all these quantities are the moduli of the corresponding vectors). The technique we put forward is based on the identification of a `fair region (FR)' in the redshift space $(r_\rmn{sp},u_\rmn{los})$. The galaxies which lie within this region will be proved to be suitable to reconstruct the total matter distribution in clusters via equation (\ref{eq:M_inf}).

We will discuss our technique using a catalogue of simulated clusters, to be able to know the overall kynematical properties of the galaxy distribution. The results of this analysis are consistent with the formulation of the SCM discussed by Cupani, Mezzetti and Mardirossian, in preparation (hereafter CMM2010) and hold within a range of values of the present-day matter density parameter ($0.2\le\Omega_{\rmn M,0}\le 0.4$, for a spatially-flat Universe). Our mass estimation technique relies only on the observable kinematic properties of cluster member galaxies, and it is therefore suitable to be applied to observed clusters. As an example, we will discuss the results of our technique when applied to a subset of clusters from the \emph{Cluster Infall Regions in the Sloan Digital Sky Survey} (CIRS) Catalogue (hereafter RD \citealt{RD}). 

In section \ref{sec:data}, we describe the simulated data catalogue we use for our analysis. In section \ref{sec:stats}, we discuss thoroughly the statistics of the member galaxy velocities. We first provide an empirical definition of the FR and then we justify this definition using the SCM. In section \ref{sec:mass_est}, we detail our mass estimation technique and its reliability. Finally in section \ref{sec:concl}, we summarize the recipe for estimating cluster masses and mass profiles and we draw the conclusion of the present work.
\section{The cluster catalogue}\label{sec:data}

We performed our analysis on the same data catalogue used by CMM08. These data were produced via a large cosmological hydrodynamical simulation of a $\Lambda$ cold dark matter ($\Lambda$CDM) Universe run by \cite{Borgal}, while the catalogue of objects (clusters and galaxies) was extracted by \cite{Bivial}. We refer to these papers for the details. 

\begin{table}
\centering
\caption[]{Cosmological parameters of the simulation, compared with the mean values (with $1\sigma$ uncertainty) obtained from the \emph{WMAP}, BAO, and SN measurements \citep[see text]{Kal}.}
\begin{tabular}{lcc}
\hline
& Simulation & WMAP+BAO+SN\\
\hline\vspace{-12pt}\\
$\sigma_8$ & $0.8$ & $0.812\pm 0.026$\vspace{2pt}\\ 
$h$ & $0.7$ & $0.705\pm 0.013$\vspace{2pt}\\ 
$\Omega_{\rmn M,0}h^2$ & $0.147$ & $0.1358_{-0.0036}^{+0.0037}$\vspace{2pt}\\
$\Omega_{\rmn{bar},0}$ & $0.04$ & $0.0456\pm 0.0015$\vspace{2pt}\\
\hline
\end{tabular}
\label{tab:cosm}
\end{table}

The simulation was run assuming a flat $\Lambda$CDM cosmology with $\Omega_{\rmn M}+\Omega_\Lambda=1 $, where $\Omega_{\rmn M}$ is the matter density parameter and $\Omega_\Lambda$ is the cosmological constant density parameter. The adopted values of the main parameters (namely the normalization of the power spectrum $\sigma_8$, the Hubble constant $h$, the present-day matter density parameter $\Omega_{\rmn M,0}$ and the present-day density parameter of baryonic matter $\Omega_{\rmn{bar},0}$) are listed in Table \ref{tab:cosm}. In the same table we also listed the observed values of these parameters, obtained by combining the last measurements of the \emph{Wilkinson Microwave Anisotropy Probe} (\emph{WMAP}) with the distance measurements from the baryonic acoustic oscillations (BAO) and from the Type Ia supernovae (SN) \citep{Kal}. The two sets of values in Table \ref{tab:cosm} evidence a quite good agreement.

\begin{figure}
\centering
\includegraphics[width=8cm]{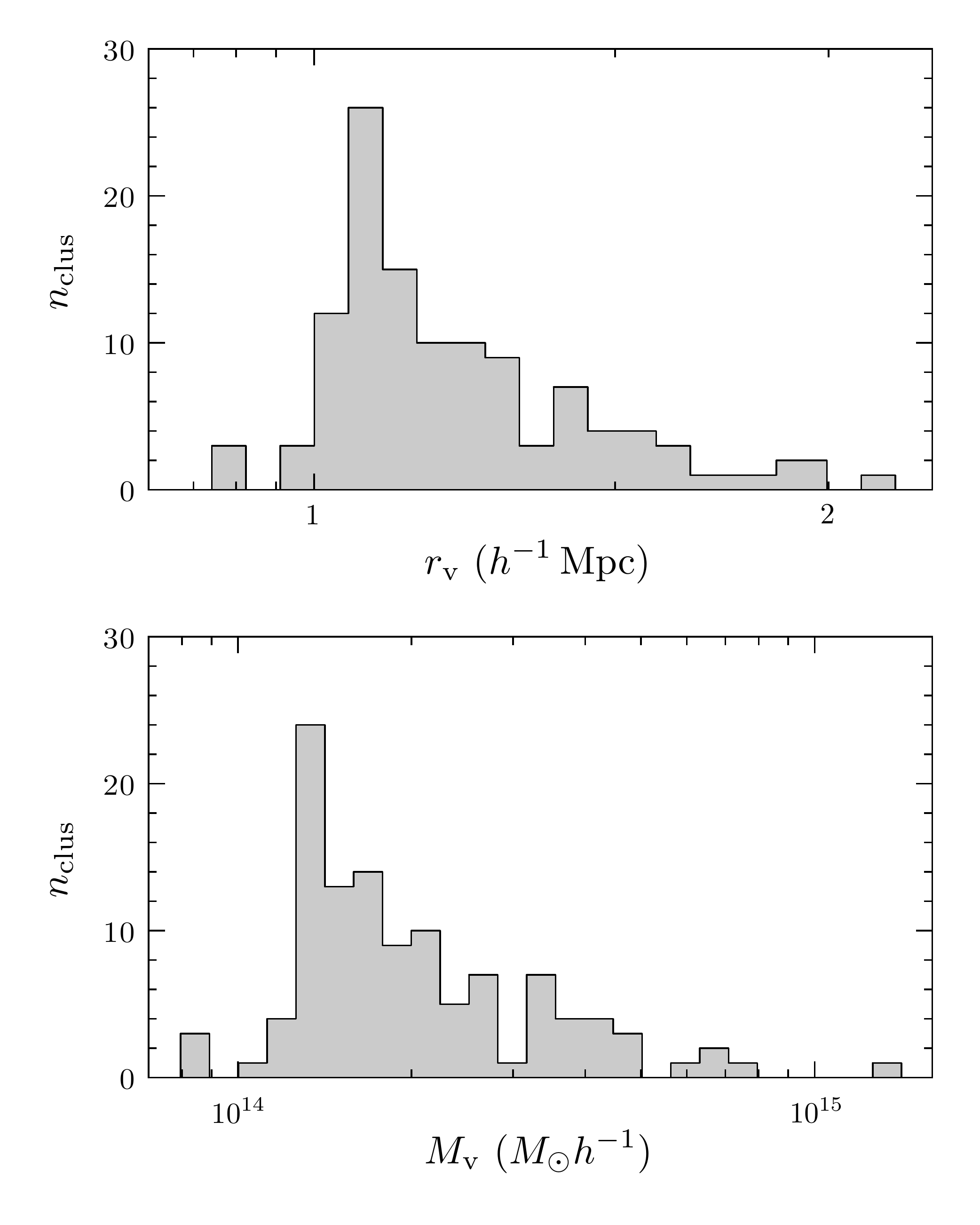}
\caption[]{Statistics of cluster sizes in the simulated data catalogue. Upper panel: frequency distribution of the virialization radii $r_\rmn v$; lower panel: frequency distribution of the virialization masses $M_\rmn v$. $n_\rmn {clus}$ is the number of clusters per frequency bin.}
\label{fig:fdistr_char}
\end{figure}

The catalogue we use \citep{Bivial} contains 114 clusters and 9631 galaxies. The clusters differ in size, with virialization radii $r_\rmn v$ in the range $r_\rmn v =(0.9\div 2.2)h^{-1}\textrm{ Mpc}$, virialization masses $M_\rmn v$ in the range $(8.0\times10^{13}\div 1.3\times 10^{15})\,h^{-1}M_\odot$, and number of member galaxies $n_\rmn{gal}$ between $17$ and $403$. The detailed frequency distributions of $r_\rmn v$ and $M_\rmn v$ are shown in Fig.~\ref{fig:fdistr_char}.

The analysis of dynamics in clusters was performed using the cluster catalogue in two ways: superimposing all clusters into a synthetic object, and studying all clusters one by one. The former approach is needed to increase the statistical significance of the results, as pointed out by \cite{VH}, while the latter approach is used to test the accuracy of the results when they are applied to single-cluster analysis. Throughout the discussion, we identify the clusters with the subscript $i$ and the galaxies with the subscript $g$.

We took into account the particles [dark matter (DM), gas, and stellar particles] of the simulation by \cite{Borgal} and the member galaxies subsequently identified by \cite{Bivial}. The particles were grouped into concentric shells in each cluster $i$, using the value of $r_{\rmn v;i}$ as normalization radius, while the member galaxies were attributed to shells according to their radial distance from the cluster centre. The shell subdivision is needed to make comparable objects of different scale. The shells were defined using a logarithmical spacing covering the whole radial extent from the virialization core to the far outskirts of clusters. We considered in our analysis the dynamically relevant quantities obtained from the simulation, namely the mass of the particles, $m$, and the six phase-space coordinates of both particles and galaxies $(x,y,z,v_x,v_y,v_z)$. The position coordinates $x$, $y$ and $z$ are referred to the cluster centre, while the velocity coordinates $v_x$, $v_y$ and $v_z$ are peculiar velocities corrected for the motion of the cluster centre. We also used the simulated data as `mock observations', i. e. restricting to only two out of six phase-space coordinates: we adopted the $x$ direction as the line of sight, and defined the sky-plane distance as $r_\rmn{sp}\equiv\sqrt{y^2+z^2}$ and the line-of-sight proper velocity as $u_\rmn{los}\equiv v_x+H_0 x$. The masses of the shells were computed by integrating the masses of all the enclosed particles. The velocities of the shells are the averages of the velocities of all the enclosed galaxies.
\section{Analysis of the redshift space}\label{sec:stats} 

\begin{figure}
\centering
\includegraphics[width=8cm]{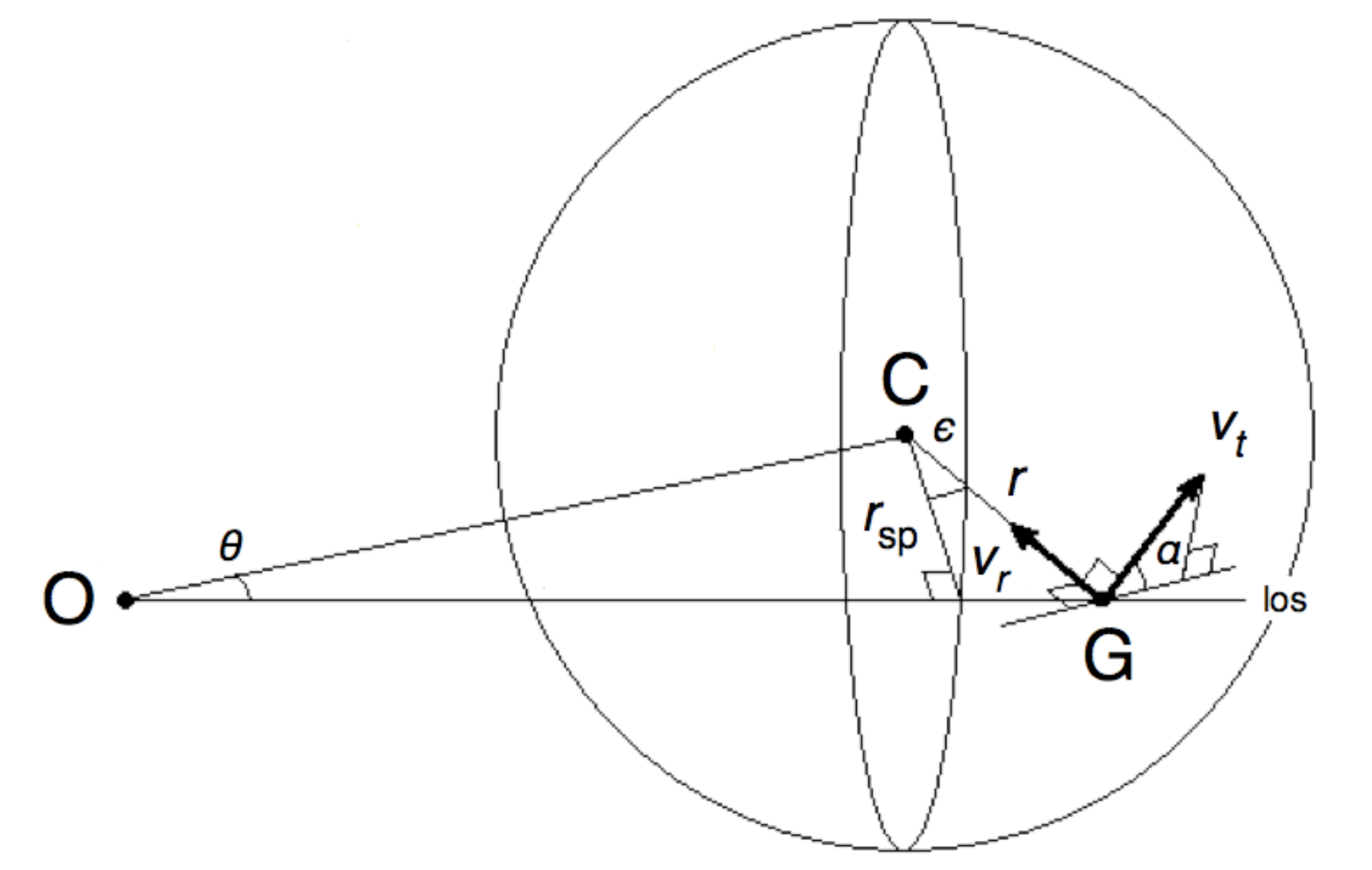}
\caption[]{Galaxy kinematics in clusters: C is the cluster centre, G is an infalling galaxy and O is a distant observer.}
\label{fig:geom}
\end{figure}

Fig.~\ref{fig:geom} represents a cluster C with an infalling galaxy G as seen by a distant observer O; in the same Fig.~one can see the angles $\epsilon$ and $\alpha$ used in the following relations. The total peculiar velocity vector $\bmath v_g$ of the galaxy is split into two ortogonal components, the radial velocity vector $\bmath v_{r;g}$ and the tangential velocity vector $\bmath v_{t;g}$: $\bmath v_g=\bmath v_{r;g}+\bmath v_{t;g}$. Within the widely accepted assumptions of spherical collapse model, the overall infall motion of galaxies can be considered as isotropic:
\be\label{eq:vr_vt_vec}
\langle\bmath v_{r;g}(r)\rangle=\bmath v_\rmn{inf}(r),\quad\langle\bmath v_{t;g}(r)\rangle=0.
\ee
Angle brackets denote the average over all galaxies located at a given radial distance $r$ from the cluster centre. It is worth noting that $\langle\bmath v_{t;g}\rangle=0$ does not generally imply $\langle v_{t;g}\rangle=0$; on the contrary, a nonzero tangential velocity modulus is expected in most of cases. In fact, as can be seen from the simulations and from the SCM (CMM08; CMM10; see also section \ref{subsec:FR}),
\be\label{eq:vr_vt}
v_\rmn{inf}\simeq\langle v_{t;g}\rangle.
\ee
The existence of random tangential motions does not conflict with the presence of an overall infall motion of galaxies, since all non-radial contribution are averagely suppressed due to isotropy. Hereinafter, we will take into account only the moduli of the velocity components.

The relation between the phase-space coordinates $r$, $v_r$ and $v_t$ and the observational redshift-space coordinates $r_\rmn{sp}$, $u_\rmn{los}$ is obtained from Fig.~\ref{fig:geom} using trigonometry. Since $u_{\rmn{los};g}$ is a proper velocity, the radial Hubble flow term $H_0 r$ must be taken into account. Therefore,
\be\label{eq:rsp}
r_{\rmn{sp};g}=r_g\cos\epsilon_g,
\ee
\be\label{eq:vlos}
u_{\rmn{los};g}=(H_0 r_g-v_{r;g})\sin\epsilon_g+v_{t;g}\cos\alpha_g\cos\epsilon_g.
\ee
Equations (\ref{eq:rsp}) and (\ref{eq:vlos}) form an undetermined system and cannot be used to compute $r_g$ and $v_{r;g}$ when only $r_{\rmn{sp};g}$ and $u_{\rmn{los};g}$ are known. To overcome this issue, we focus on the subset of all galaxies having $r_{\rmn{sp};g}\simeq r_g$ and satisfying the condition $\langle|u_{\rmn{los};g}|\rangle\simeq v_\rmn{inf}$ (see below). We will refer to these galaxies as `tracking galaxies' (hereinafter TGs or the TG subset), since they are the best candidates to estimate the overdensity and the mass profile using only the observed redshift-space coordinates. By definition, the TGs have
\be\label{eq:omegainf}
\omega_\rmn{inf}(r)\simeq\f{\langle |u_{\rmn{los};g}(r_\rmn{sp})|\rangle}{H_0r_\rmn{sp}}.
\ee
Substituting equations (\ref{eq:rsp}) and (\ref{eq:omegainf}) into equation (\ref{eq:M_inf}), one can compute the cluster mass profile using only the observable properties of the TGs:
\begin{eqnarray}\label{eq:M_inf_TG}
M(r)&\simeq& M_\rmn{inf}\big[r_\rmn{sp},\langle |u_{\rmn{los};g}(r_\rmn{sp})|\rangle\big]\nonumber\\
&\equiv&\f{4}{3}\pi\rho_{\rmn{cr},0}\Omega_{\rmn M,0}r_\rmn{sp}^3\!\left\{\!1+F^{-1}\!\!\left[\Omega_{\rmn M,0}^{-0.6}\f{\langle |u_{\rmn{los};g}(r_\rmn{sp})|\rangle}{H_0r_\rmn{sp}}\!\right]\!\right\}\!.
\end{eqnarray}

The statistical properties of the TG subset are essential to determine how the infall velocity approach can be used to estimate the cluster mass profiles. TGs are not randomly distributed among other galaxies, but tend to be gathered in an identifiable region of the redshift space. By definition, the TGs have $\epsilon_g\simeq 0$, which substituted into equation (\ref{eq:vlos}) yields $u_{\rmn{los};g}\simeq v_{t;g}\cos\alpha_g$. Combining this result with equation (\ref{eq:vr_vt}), one obtains that $\langle\alpha_g\rangle\simeq 0$. Therefore, at any given sky-plane distance, the TGs satisfy the condition
\be\label{eq:vlos_vr_vt}
\langle |u_{\rmn{los};g}(r_\rmn{sp})|\rangle\simeq v_\rmn{inf}\simeq\langle v_{t;g}(r_\rmn{sp})\rangle.
\ee
Equation (\ref{eq:vlos_vr_vt}) suggests that TGs are unlikely to be found in the region of the redshift space where $u_\rmn{los}\simeq 0$, since both $v_r$ and $v_t$ are generally non-zero.

\begin{figure}
\centering
\includegraphics[width=8cm]{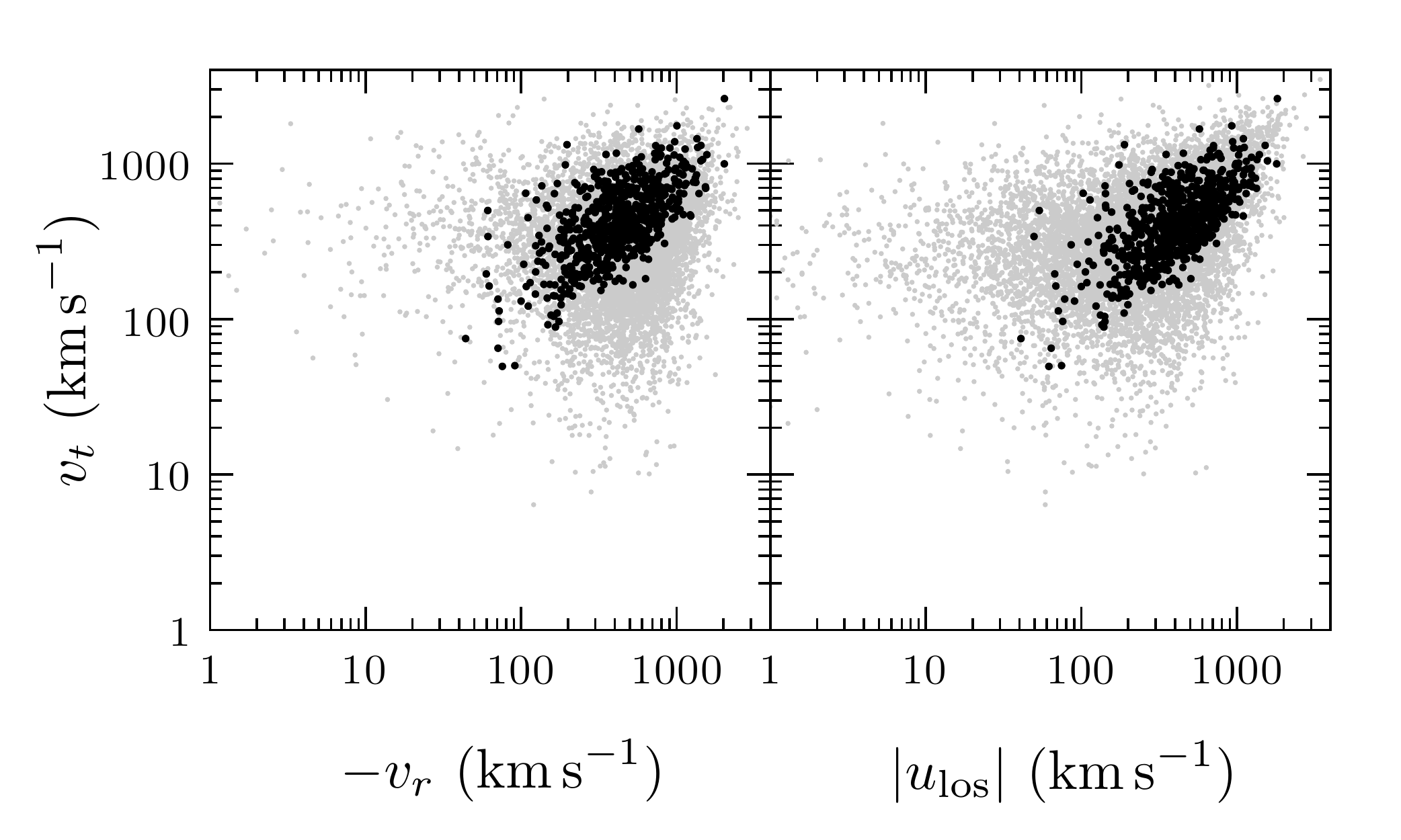}
\caption[]{Dependence between the velocity components of galaxies. The grey dots are all galaxies extracted from the simulation. The black dots are the TGs, defined as in equations (\ref{eq:rsp_cond}) and in (\ref{eq:vlos_cond}) with $\eta_\rmn{TG}^\star=0.2$.}
\label{fig:vt-vrvlos}
\end{figure}

These considerations are supported by the analysis of the simulated data of our catalogue. We represented in Fig.~\ref{fig:vt-vrvlos}, the dependence between $v_{r;g}$ and $v_{t;g}$ (left-hand column) and between $|u_{\rmn{los};g}|$ and $v_{t;g}$ (right-hand column) as extracted from our data catalogue. The galaxies are represented with grey dots, while the TGs are represented as black dots. The TGs were operatively defined as satisfying the conditions
\be\label{eq:rsp_cond}
\f{r_{\rmn{sp};g}}{r_g}=\cos\epsilon_g\ge 1-\eta_\rmn{TG}^\star,
\ee
\be\label{eq:vlos_cond}
1-\eta_\rmn{TG}^\star\le\f{|u_{\rmn{los};g}|}{v_{r;g}}\le 1+\eta_\rmn{TG}^\star.
\ee
where the parameter $\eta_\rmn{TG}^\star=0.2$ is used to mimic a quite small discrepancy around $0$. The TG subset selected with this choice contains $579$ galaxies. 

\begin{figure}
\centering
\includegraphics[width=8cm]{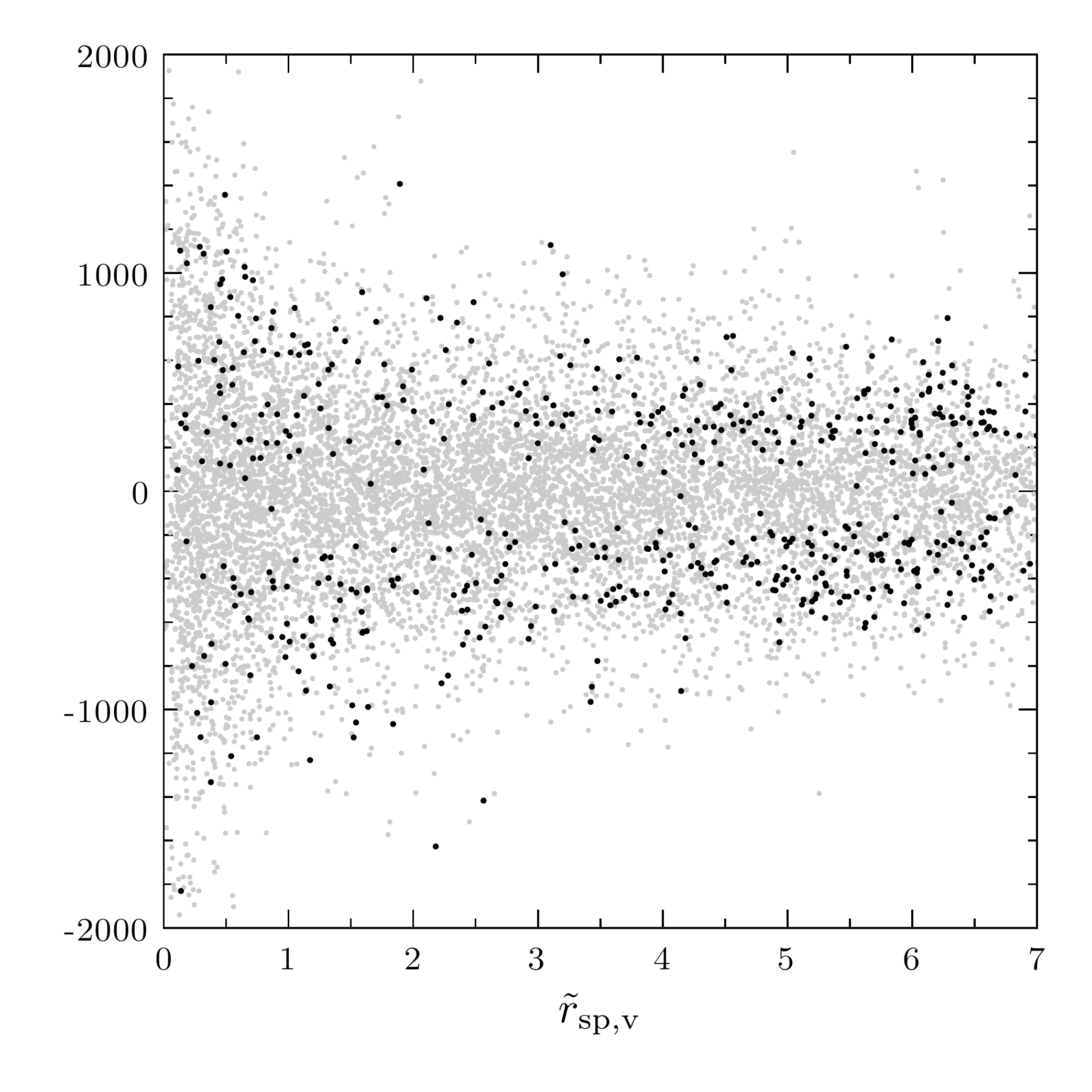}
\caption[]{Redshift-space distribution of galaxies. The grey dots are all galaxies extracted from the simulation. The black dots are the TGs, defined as in equations (\ref{eq:rsp_cond}) and (\ref{eq:vlos_cond}) with $\eta_\rmn{TG}^\star=0.2$.}
\label{fig:rs}
\end{figure}

The redshift-space distribution of galaxies extracted from our data catalogue is shown in Fig.~\ref{fig:rs}. Also in this case, the galaxies are represented with grey dots, while the TGs are represented as black dots. The abscissa is the sky-plane distance in units of the virialization radius $\tilde r_{\rmn{sp,v}}\equiv r_{\rmn{sp}}/r_{\rmn v}$. Normalization to $r_{\rmn v}$ is required to make comparable clusters of different size. The ordinate is the line-of-sight velocity $u_{\rmn{los};g}$. The resulting distribution shows a well-defined trumpet shape, which is typical of caustic surfaces and reaches its minimum amplitude at the turnaround (see e.g. \citealt{Oal,RG}). The TGs are concentrated in two narrow bands of the redshift space, and are almost absent in the region where $u_\rmn{los}\simeq 0$.

\subsection{The Fair Region of the redshift space}\label{subsec:FR}

We can roughly define as `FR' the region of the redshift space where the TGs are mostly concentrated. In fact, an effective definiton of the FR is required to estimate the mass profile of clusters with the member galaxies via the infall velocity approach. We will obtain such definition combining a statistical analysis of the simulated data catalogue with the theoretical predictions provided by the SCM (CMM10).

\begin{figure}
\centering
\includegraphics[width=8cm]{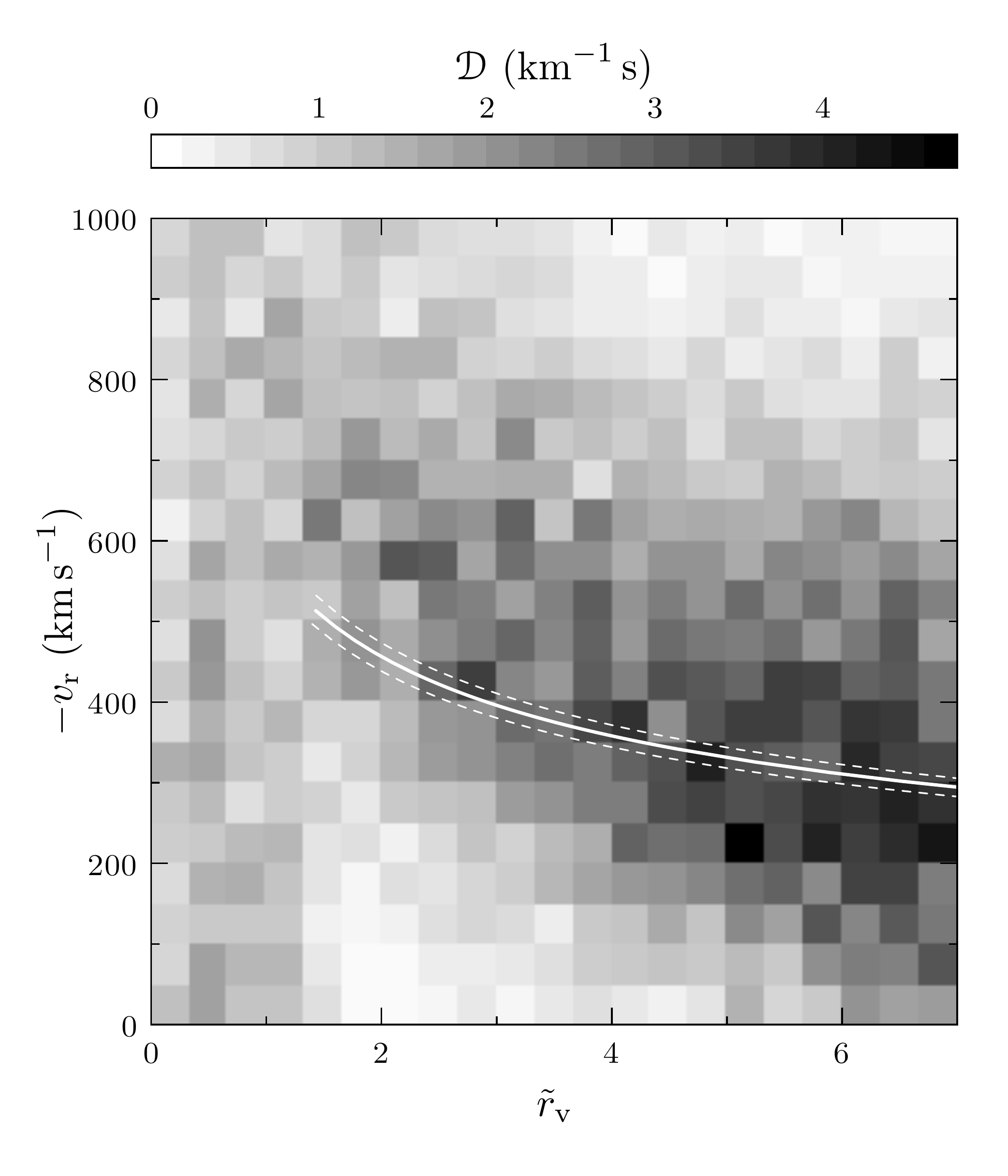}
\caption[]{Distribution of galaxy radial velocities, compared with the prediction of the SCM (from CMM10). The grey-scale represents the number density of galaxies for unit surface. The lines are the theoretical radial velocity profile computed with the SCM, assuming $\Omega_\rmn M=0.3$ (thick solid line) ans $\Omega_\rmn M$ equal to $0.2$ and $0.4$ (lower and upper dashed line, respectively).}
\label{fig:vr-r_det}
\end{figure}

\begin{figure}
\centering
\includegraphics[width=8cm]{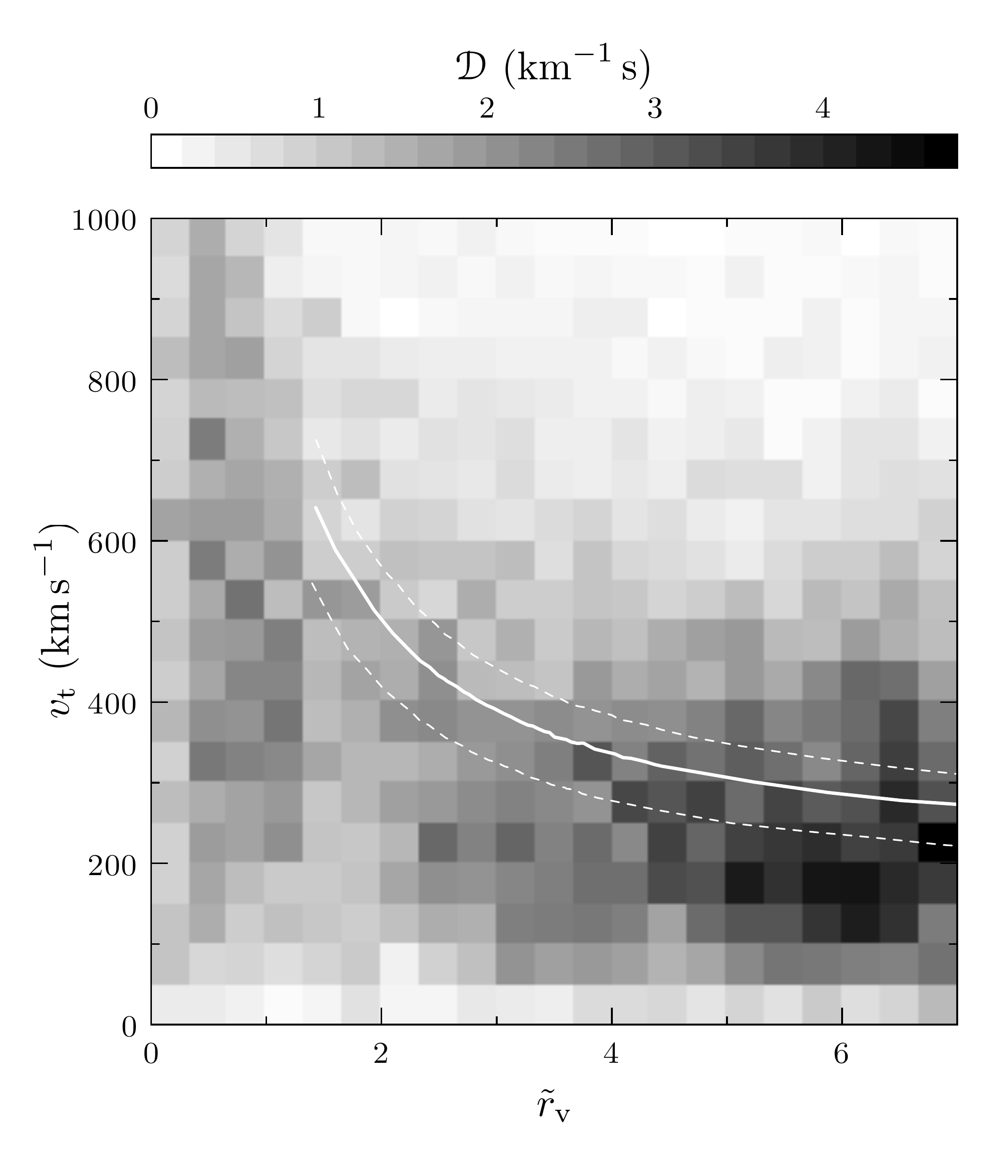}
\caption[]{Distribution of galaxy tangential velocities, compared with the prediction of the SCM (from CMM10). The grey-scale represents the number density of galaxies for unit surface. The lines are the theoretical tangential velocity profile computed with the SCM, assuming $\Omega_\rmn M=0.3$ (thick solid line) ans $\Omega_\rmn M$ equal to $0.2$ and $0.4$ (lower and upper dashed line, respectively).}
\label{fig:vt-r_det}
\end{figure}

\begin{figure}
\centering
\includegraphics[width=8cm]{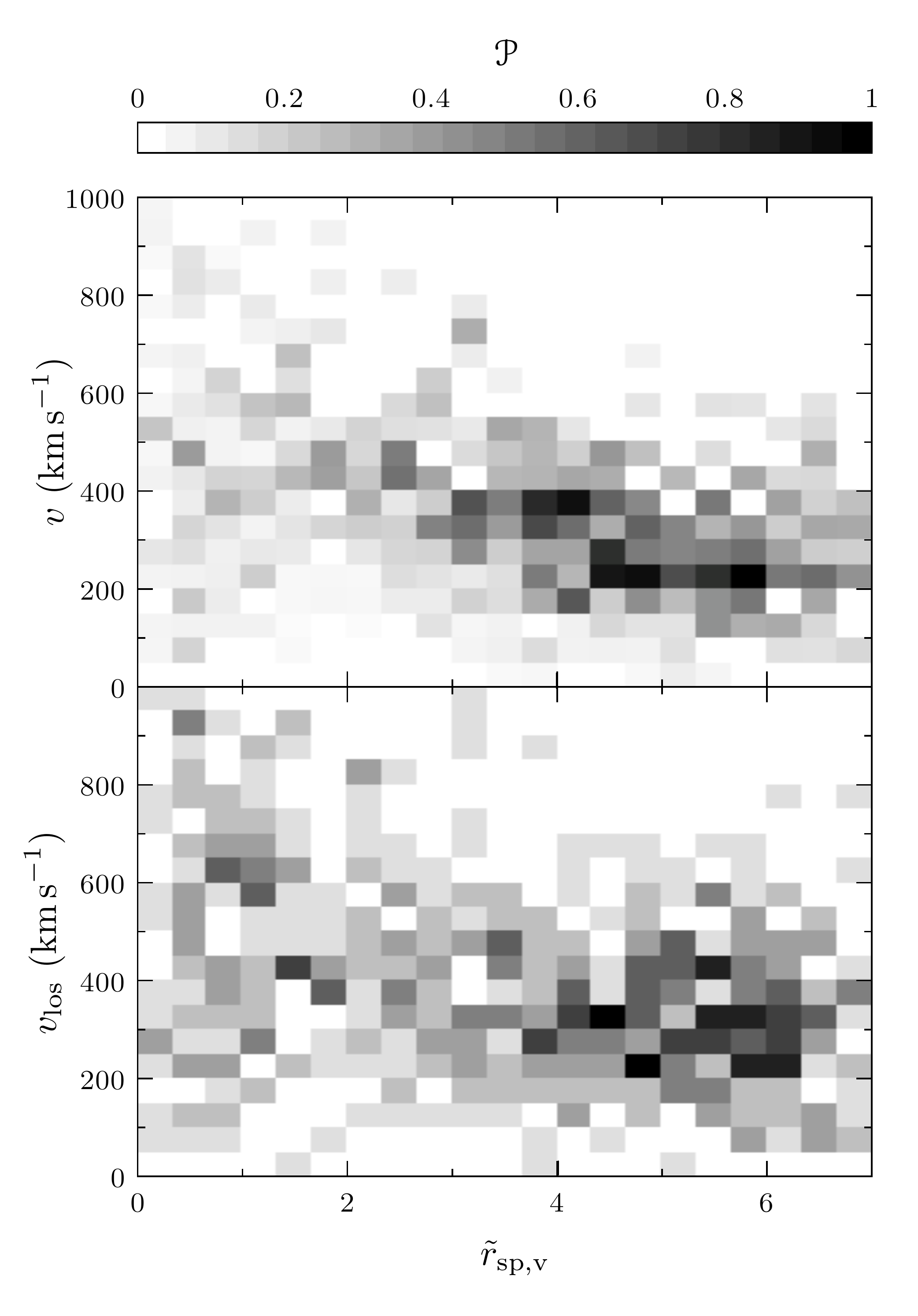}
\caption[]{Statistical identification of the FR in the redshift space. The upper panel is the expected distribution of TGs in the redshift space, computed as the joint probability of the distributions in Figs \ref{fig:vr-r_det} and \ref{fig:vt-r_det}. The lower panel is the distribution of TGs in the redshift space extracted from the simulation. The grey-scale represents the probability density of galaxies for unit surface.}
\label{fig:FR_TG}
\end{figure}

According to the condition of equation (\ref{eq:vlos_vr_vt}), the FR is defined as the region where the mean peculiar radial velocity is almost equal to the mean tangential velocity. The $v_r$ and $v_t$ distribution of our galaxies along $r$ is represented in Figs \ref{fig:vr-r_det} and \ref{fig:vt-r_det}, respectively. To determine the grey-scale, we superimposed an orthogonal grid on each plane and counted the galaxies in each cell of the grid, in order to produce a two-dimensional histogram. The grey-scale represents the number density of galaxies for unit surface $\mathcal D$. Both distribution are interpolated with the SCM prediction for $\Omega_{\rmn M,0}=0.3$ (thick solid line) and for $\Omega_{\rmn M,0}=0.2$ and $\Omega_{\rmn M,0}=0.4$ (lower and upper dashed line, respectively). Despite the notable dispersion due to the variance among clusters (see CMM08), the SCM predicts quite well the main profile of both $v_r$ and $v_t$. The dependence on $\Omega_{\rmn M,0}$ is overshadowed by the data dispersion and is generally negligible. To determine the location of the FR, we computed the joint probability for galaxies to be located in the same grid cell of Figs \ref{fig:vr-r_det} and \ref{fig:vt-r_det}. Fig.~\ref{fig:FR_TG} represents the distribution obtained by this procedure (upper panel), compared to the distribution of TGs from the simulated data catalogue. Here the grey-scale represents the probability parameter $\mathcal P$, obtained by normalizing $\mathcal D$ to unity over the whole plane. The two panels show a remarkable agreement, indicating that the condition in equation (\ref{eq:vlos_vr_vt}) is suitable to identify the FR.

\begin{figure}
\centering
\includegraphics[width=8cm]{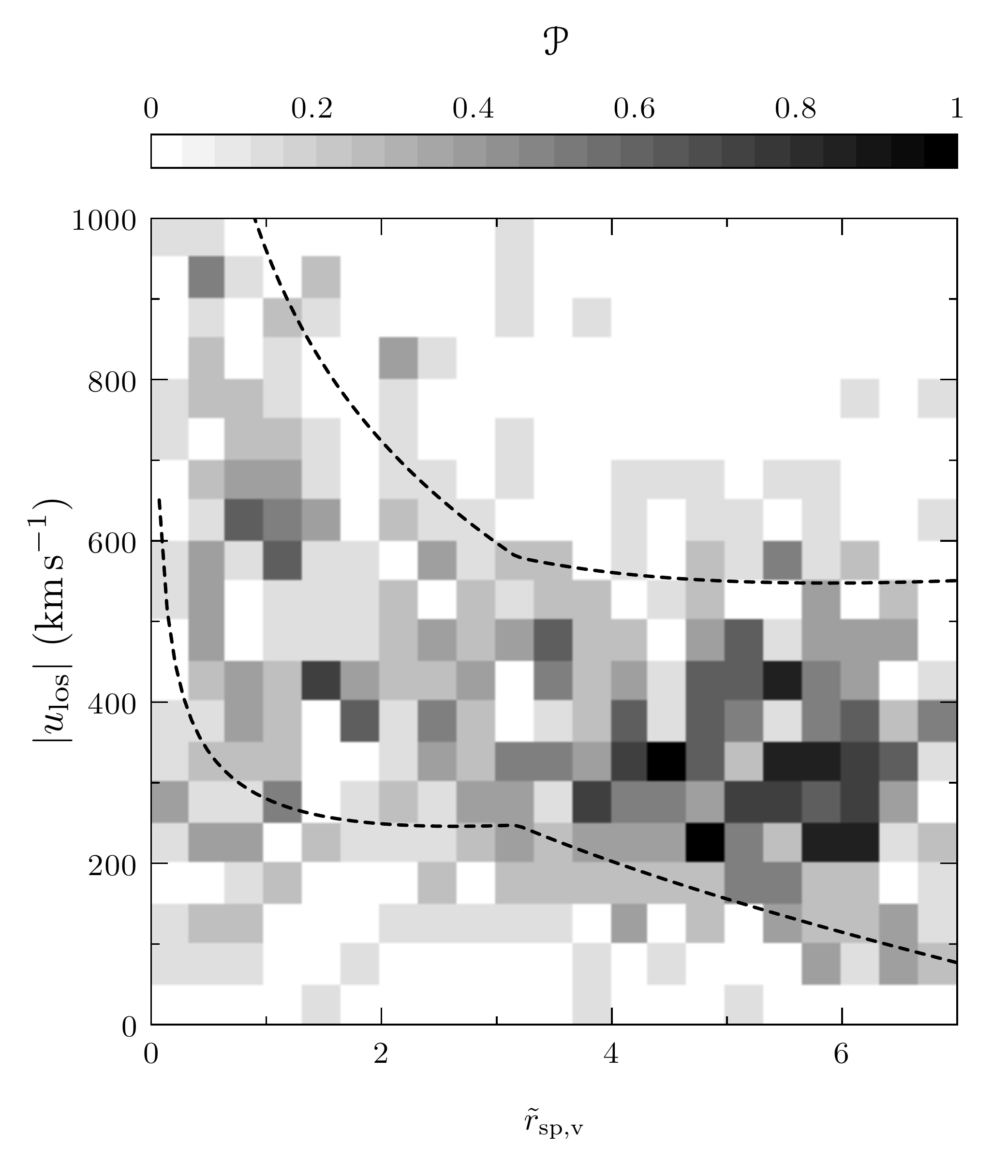}
\caption[]{Theoretical identification of the FR in the redshift space. The grey-scale represents the probability density of TGs for unit surface (same as in the lower panel of Fig.~\ref{fig:FR_TG}). The dashed lines are the boundaries of the FR obtained through equation (\ref{eq:Kv_cond}) and equation (\ref{eq:epsilon_alpha_cond}) with $K_{v;\rmn{FR}}^{*}=1.5$ and $\eta_{\rmn{FR}}^{*}=0.2$.}
\label{fig:FR}
\end{figure}

The empirical definition of the FR is theoretically confirmed by the SCM results (CMM10). In a spherically symmetric scenario, the FR is theoretically defined as the region where the parameters $\Delta r\equiv 1-r_\rmn{sp}/r$ and $\Delta v\equiv 1-u_\rmn{los}/v_r$ are minimized, in agreement with the definition of the TGs. Considering equations (\ref{eq:rsp}) and (\ref{eq:vlos}), we can write
\be\label{eq:Deltar}
\Delta r=1-\cos\epsilon,
\ee
\be\label{eq:Deltav}
\Delta v=1-\left(\f{H_0 r}{v_r}-1\right)\sin\epsilon+K_v\cos\alpha\cos\epsilon,
\ee
where $K_v\equiv v_t/v_r$. Equations (\ref{eq:Deltar}) and (\ref{eq:Deltav}) describe how the infall velocity profile predicted by the SCM is biased due to projection on to the sky plane. This bias depends both on the variance in galaxy dynamics (parametrized by $K_v$ and $\alpha$) and on the galaxy displacement with respect to the sky plane (parametrized by $\epsilon$). The extension of the FR can be therefore determined by taking the extremal values of $\Delta r$ and $\Delta v$ obtained by varying $K_v$ and $\epsilon$ and $\alpha$ around their mean values. In fact, CMM10 showed that the mean value of $K_v$ is quite close to unity in the non equilibrium region for $0.2\le\Omega_{\rmn M,0}\le 0.4$, in agreement with equation (\ref{eq:vr_vt}). In particular, if $\Omega_{\rmn M,0}=0.3$, $\langle K_v(\tilde r_v)\rangle\simeq 0.95$ for $\tilde r_v\ge 3$. So, in order to operatively define the FR, we used the following conditions: 
\be\label{eq:Kv_cond}
\f{0.95}{K_{v;\rmn{FR}}^{*}}\le K_v\le 0.95 K_{v;\rmn{FR}}^{*},\quad K_{v;\rmn{FR}}^{*}=1.5;
\ee
\be\label{eq:epsilon_alpha_cond}
-\eta_{\rmn{FR}}^*\le\epsilon,\alpha\le\eta_{\rmn{FR}}^*,\quad\eta_{\rmn{FR}}^*=0.2.
\ee
The parameter $K_{v;\rmn{FR}}^*$ mimics the variance of $K_v$ around its mean value $0.95$, while the parameter $\eta_{v;\rmn{FR}}$ mimics a small discrepancy of $\epsilon$, and $\alpha$ around their mean value $0$. The boundaries of the FR obtained through equation (\ref{eq:Kv_cond}) and (\ref{eq:epsilon_alpha_cond}) are shown as dashed lines in Fig.~\ref{fig:FR}, superimposed to the grey-scale distribution of TGs (same as in the lower panel of Fig.~\ref{fig:FR_TG}). The agreement between the theoretical prediction and the data distribution of our catalogue is remarkable. This result shows that the FR, whose existence was originally inferred from the distribution of the TGs in the redshift space, is due to the projection of the infall velocity profile onto the sky plane. The SCM is shown to work for $0.2\le\Omega_{\rmn M,0}\le 0.4$; this allows to identify the FR in different cosmologies. 
\section{Mass estimation}\label{sec:mass_est}

The definition FR in the redshift space provides a way to select the galaxies which are suitable to reconstruct the total matter distribution of clusters. The galaxies lying within the FR will be referred to as `Fair galaxies' (hereinafter FGs or the FG subset). In this section, we will prove that the infall velocity approach is suitable to estimate the mass profiles of clusters using the FG subset. 

Our technique can be outlined as follows:
\begin{enumerate}
\p identification of the FG subset via the identification of the FR provided in Section \ref{subsec:FR};
\p detection of the mass profiles, using the redshift-space coordinates of the FGs.
\end{enumerate}
We used the simulated data catalogue to test the reliability of such approach, comparing the estimated mass profiles obtained with our technique with the actual mass profiles of the simulated clusters. The simulated data were transformed into `mock observations' as described in Section \ref{sec:data}. All galaxies lying within the dashed lines in Fig.~\ref{fig:FR} were regarded as FGs, and their observable redshift-space coordinates $r_\rmn{sp}$ and $u_\rmn{los}$ were used instead of the unknown quantities $r$ and $v_r$ to reconstruct the infall pattern and to estimate the mass profile of clusters.

Before applying the mass estimation technique to the FG subset, we took into account the effect of projection of the galaxy radial positions. In fact, while the TGs approximately lie on the sky plane, the FGs are in principle distributed all along the line of sight. Therefore, using $r_{\rmn{sp};g}$ instead of $r_g$ for the FG subset would introduce a bias in the estimation of the mass along the radial coordinate. To minimize this bias, we introduce a `guess' radial distance, defined as follows:
\be
r_{\rmn{G};g}\equiv\f{r_{\rmn{sp};g}}{\mathcal R(r_{\rmn{sp};g})},
\ee
where $\mathcal R(r_{\rmn{sp}})$ is the mean ratio between $r_{\rmn{sp};g}$ and $r$ at a given sky-plane distance $r_\rmn{sp}$. The guess radial distance $r_{\rmn{G};g}$ will be used hereinafter as a replacement of $r_{\rmn{sp};g}$ when handling the FG subset. If the FG distribution is spherically symmetric, $\mathcal R$ is obtained as
\be\label{eq:R}
\mathcal R(r_\rmn{sp})=\f{\displaystyle\int_{0}^{\sqrt{r_\rmn{cut}^2-r_\rmn{sp}^2}}{\f{\nu\left(\sqrt{r_\rmn{sp}^2+r_\rmn{los}^2}\right) r_\rmn{sp}}{\sqrt{r_\rmn{sp}^2+r_\rmn{los}^2}}\rmn dr_\rmn{los}}}{\displaystyle\int_{0}^{\sqrt{r_\rmn{cut}^2-r_\rmn{sp}^2}}{\nu\left(\sqrt{r_\rmn{sp}^2+r_\rmn{los}^2}\right)\rmn dr_\rmn{los}}},
\ee
where $r_\rmn{los}$ is the line-of-sight projection of the radial coordinate, $r_\rmn{cut}$ is a cut radius, and $\nu(r)$ is the number density of galaxies. Using the SCM predictions for $\nu(r)$ and assuming $r_\rmn{cut}=7 r_v$ (since generally we have no appreciable data beyond this distance), we approximated equation (\ref{eq:R}) with a linear fitting algorithm as $\mathcal R(\tilde r_\rmn{sp})=5.5\times10^{-2}\tilde r_\rmn{sp}+0.6$. This fit is in agreement with the data distribution and is used to correct the radial position of FGs in our simulated data catalogue.

\begin{figure}
\centering
\includegraphics[width=8cm]{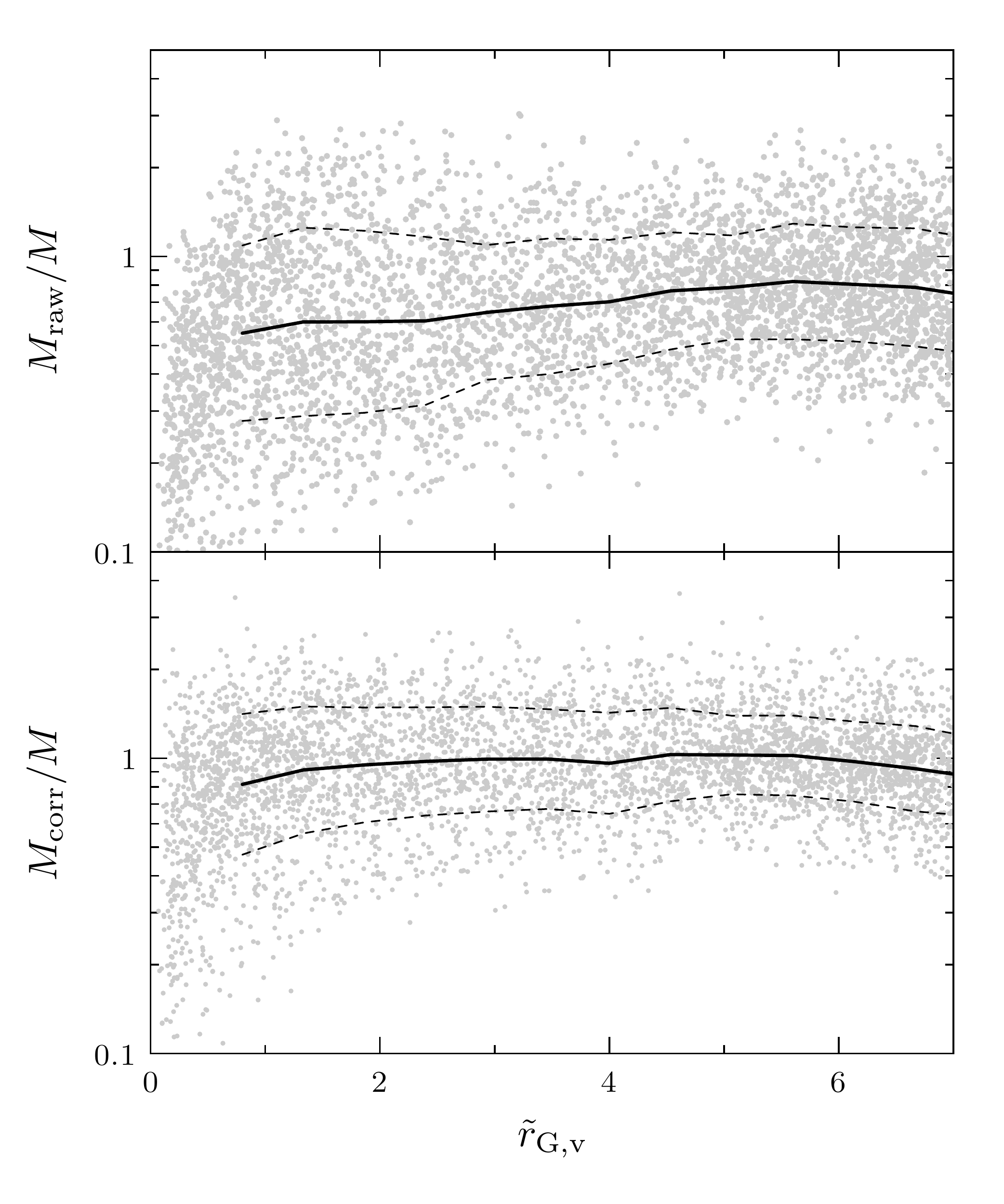}
\caption[]{Reliability of the raw mass estimate and of the corrected mass estimate using the FG subset. The grey dots in the upper panel correspond to the values of $M_{\rmn{raw};g}/M_g$, while those in the lower panel correspond to the values of $M_{\rmn{corr};g}/M_g$, both represented as a function of $r_{\rmn{G,v};g}$. The average profile and the $1\sigma$ uncertainty region of the distribution along the radial coordinate are shown in both panels with thick solid lines and dashed lines, respectively.
}
\label{fig:M_r}
\end{figure}

\begin{table*}
\centering
\caption[]{Accuracy of mass estimation technique, evaluated by taking into account a mimicked observational error $\Delta u_\rmn{los}^*$ in the line-of-sight velocity and the radial interval of aperture in observation.}
\begin{tabular}{ccccccc}
\hline
$\Delta u_\rmn{los}^\star$ $(\textrm{km h}^{-1})$& interval & $n_\rmn{gal}$ & $\mu_M$ & $\sigma_M$ & $\mu_M\sigma_M$ & $\mu_M/\sigma_M$\\
\hline
& $[1.0,7.0ö]$ & $3693$ & $0.97$ & $1.46$ & $1.42$ & $0.67$\\
$0$ & $[1.0,3.5]$ & $1383$ & $0.96$ & $1.56$ & $1.49$ & $0.62$\\ 
& $[1.0,2.2]$ & $743$ & $0.93$ & $1.59$ & $1.48$ & $0.58$\vspace{0.1cm}\\
& $[1.0,7.0ö]$ & $3746$ & $0.99$ & $1.48$ & $1.46$ & $0.67$\\
$100$ & $[1.0,3.5]$ & $1421$ & $0.96$ & $1.58$ & $1.52$ & $0.61$\\ 
& $[1.0,2.2]$ & $752$ & $0.92$ & $1.62$ & $1.49$ & $0.57$\vspace{0.1cm}\\
& $[1.0,7.0ö]$ & $3840$ & $1.02$ & $1.47$ & $1.50$ & $0.70$\\
$200$ & $[1.0,3.5]$ & $1504$ & $1.01$ & $1.58$ & $1.59$ & $0.64$\\ 
& $[1.0,2.2]$ & $798$ & $0.97$ & $1.60$ & $1.57$ & $0.60$\vspace{0.1cm}\\
& $[1.0,7.0ö]$ & $3761$ & $1.07$ & $1.51$ & $1.61$ & $0.71$\\
$300$ & $[1.0,3.5]$ & $1488$ & $1.05$ & $1.62$ & $1.69$ & $0.65$\\ 
& $[1.0,2.2]$ & $802$ & $1.00$ & $1.67$ & $1.67$ & $0.60$\\
\hline
\end{tabular}

\emph{Note:} $\mu_M$ and $\sigma_M$ are the moments of the distributions in the lower panel of Fig.~\ref{fig:M_r}, computed for different intervals of radial aperture.
\label{tab:acc}
\end{table*}

The first raw estimate of the cluster mass profiles of is obtained by substituting the values of $r_{\rmn{G};g}$ and $u_{\rmn{los};g}$ into equation (\ref{eq:M_inf_TG}):
\be
M_{\rmn{raw};g}\equiv M_\rmn{inf}(r_{\rmn{G};g},|u_\rmn{los};g|).
\ee
The function $F^{-1}$ was defined, according to CMM08, as the inverse of the Meiksin approximation \citep{VD}:
\be
F(\delta)=\f{\delta}{3}\left(1+\f{\delta}{3}\right)^{-1/2}.
\ee
The upper panel of Fig.~\ref{fig:M_r} shows the distribution obtained computing the ratio between $M_{\rmn{raw};g}$ and the real mass values $M_g$ extracted from the simulation, for each galaxy in the FG subset (grey dots). The average and the $1\sigma$ uncertainty region of the distribution along the radial coordinate is also shown (thick solid line and dashed lines, respectively). The radial coordinate is normalized to the virialization radius, $\tilde r_{\rmn{G,v};g}\equiv r_{\rmn{G};g}/r_\rmn v$. The distribution is significantly lower than $1$ in the cluster core and approach unity in the non-equilibrium region. This is due both to projection effects and to the fact that the infall velocity approach is mostly reliable in the cluster outskirts, as already pointed out by CMM08. To correct the underestimation, we approximated the average distribution in the upper panel of Fig.~\ref{fig:M_r} with a linear fitting algorithm and used it as a corrective term to compute a `guess' mass value, as follows:
\be\label{eq:Mguess}
M_{\rmn{G};g}\equiv\f{M_{\rmn{raw};g}}{\mathcal M(\tilde r_{\rmn{G,v};g})},
\ee
where $\log_{10}\mathcal M(\tilde r_{\rmn{G,v};g})=0.025\tilde r_{\rmn{G,v};g}-0.26$.

The final corrected estimate of the cluster mass profiles, $M_{\rmn{corr};g}$, was obtained by smoothing the distribution of $M_{\rmn{G};g}$ values for each cluster, taking its running median along the radial coordinate. We computed the running median within a window defined to contain approximately one-third of the galaxies in each cluster. The lower panel of Fig.~\ref{fig:M_r} shows the distribution obtained computing the ratio between $M_{\rmn{corr};g}$ and the real mass values $M_g$ extracted from the simulation, for each galaxy in the FG subset (grey dots). Also in this case, the average and the $1\sigma$ uncertainty region of the distribution along the radial coordinate is also shown (thick solid line and dashed lines, respectively). The combination of correction and running median smoothing yields very good results. The overall distribution is very close to unity from $1$ to $7$ virialization radii. The variance of the distribution decreases when moving outwards, confirming once more the reliability of the infall velocity approach in the cluster non-equilibrium region.

The overall accuracy of our technique was evaluated by computing the statistical moments of the distribution of $M_{\rmn{corr};g}/M_g$ values. We denote with $\mu_M$ and $\sigma_M$ the logarithmical mean and the logarithmical variance computed among the set of values $M_\rmn{corr}/M$:
\be
\mu_M\equiv\exp\left[\f{1}{n_\rmn{gal}}\sum_{g=1}^{n_\rmn{gal}}\big(\ln M_{\rmn{corr};g}-\ln M_g\big)\right],
\ee
\be
\sigma_M\equiv\sqrt{\exp\left[\f{1}{n_\rmn{gal}}\sum_{g=1}^{n_\rmn{gal}}\big(\ln M_{\rmn{corr};g}-\ln M_g-\mu_M\big)^2\right].}
\ee
These moments were computed under different conditions, by taking into account the effects of the observational error and of the radial aperture in cluster observation. In particular, 
\begin{enumerate}
\p we mimicked the observational error associated to the line-of-sight velocity measurement by substituting $u_{\rmn{los};g}$ with a perturbed line-of-sight velocity $\widehat u_{\rmn{los};g}$:
\be\label{eq:vlos_pert}
\widehat u_{\rmn{los};g}\equiv u_{\rmn{los};g}+\Delta u_{\rmn{los};g},\quad\Delta u_{\rmn{los};g}\equiv\rmn{rand}(0,\Delta u_\rmn{los}^\star).
\ee
Here, $\rmn{rand}(\mu,\sigma)$ is a function which generates random values having a Gaussian distribution with mean $\mu$ and variance $\sigma$. We set $\Delta u_\rmn{los}^*$ alternatively equal to $0$, $100$, $200$ and $300$ km s$^{-1}$; 
\p we defined different intervals of $\tilde r_{\rmn{G,v};g}$, namely $[1,7]$, $[1,3.5]$, and $[1,2.2]$, respectively, corresponding to the extent of uniform data coverage in our data catalogue, to the region encompassed by turnaround radius (CMM08; CMM10), and to the transition region between the cluster core and the non-equilibrium region (where the infall approach is expected to be less reliable; see CMM08). The cluster core was rejected in all cases to avoid the contamination from virialization-related phenomena, which undermine the reliability of infall approach.
\end{enumerate}

The values of $\mu_M$ and $\sigma_M$ obtained with different choices of $\Delta u_\rmn{los}^*$ and of the aperture interval are listed in Table \ref{tab:acc}. The accuracy is very good in all cases, confirming the reliability of the infall velocity approach applied to the FG subset. The variance $\sigma_M$ is only slightly affected by the observational error on the line-of-sight velocity and by the radial coverage in the galaxy distribution, confirming the robustness of the estimation.

Our technique is also suitable to compute the virialization mass and turnaround mass of clusters. $M_\rmn{corr,v}$ and $M_\rmn{corr,t}$ were obtained by interpolating the distribution of $M_{\rmn{corr};g}$ values at $\tilde r_\rmn{G,v}=1$ and at the radius where the overdensity equals the turnaround value $\delta_\rmn t=15$ (CMM08; CMM10). The interpolation was done with a linear fitting algorithm. In fact, this procedure is ineffective when the running median does not cover the radial extent from the virialization radius to the turnaround radius. We denote with $f_\rmn{fail,v}$ and $f_\rmn{fail,t}$, respectively, the fraction of clusters for which the estimation of $M_\rmn{corr,v}$ and $M_\rmn{corr,t}$ is impossible. Focusing on the remaining cases, we denote with $f_\rmn{agree,v}$ and $f_\rmn{agree,t}$, respectively, the fraction of clusters for which the estimated values are consistent with the actual values $M_\rmn v$ and $M_\rmn t$ within the uncertainty. The values of $f_\rmn{fail,v}$, $f_\rmn{agree,v}$, $f_\rmn{fail,t}$, and $f_\rmn{agree,t}$ obtained for different choices of $\Delta u_\rmn{los}^*$ are listed in Table \ref{tab:acc_MvMt}. The percentage of failure is slightly increased when the observational error in the line-of-sight velocity is introduced, as an effect of the higher noise in the galaxy velocity distribution. Nevertheless, the estimation of both $M_\rmn v$ and $M_\rmn t$ is significantly accurate even for relatively high values of $\Delta u_\rmn{los}^*$.

\begin{table}
\centering
\caption[]{Accuracy of the virialization and turnaround mass estimate, evaluated by taking into account a mimicked observational error $\Delta u_\rmn{los}^*$ in the line-of-sight velocity and the radial interval of aperture in observation.
}
\begin{tabular}{ccccc}
\hline
$\Delta u_\rmn{los}^\star$ $(\textrm{km h}^{-1})$ & $f_\rmn{fail,v}$ & $f_\rmn{agree,v}$ & $f_\rmn{fail,t}$ & $f_\rmn{agree,t}$\\
\hline
$0$ & $0.10$ & $0.90$ & $0.06$ & $0.72$\\
$100$ & $0.15$ & $0.93$ & $0.04$ & $0.69$\\
$200$ & $0.19$ & $0.91$ & $0.03$ & $0.59$\\
$300$ & $0.19$ & $0.91$ & $0.04$ & $0.54$\\
\hline
\end{tabular}

\emph{Note:} $f_\rmn{fail,v}$ and $f_\rmn{fail,t}$ are the fractions of clusters for which the estimation is impossible (computed among all clusters). $f_\rmn{agree,v}$ and $f_\rmn{agree,t}$ are the fractions of clusters for which the estimated value is in agreement with the actual value within the uncertainty (computed among the clusters for which the estimation is possible).
\label{tab:acc_MvMt}
\end{table}
\subsection{Single-cluster estimations}

Some characteristics of our mass estimation technique can be better appreciated by comparing the mass profiles estimated for single clusters separately. In this section, we will discuss in detail the results obtained for nine simulated clusters from our data catalogue and for nine observed clusters from the CIRS Catalogue by RD.

\begin{figure*}
\centering
\includegraphics[width=16cm]{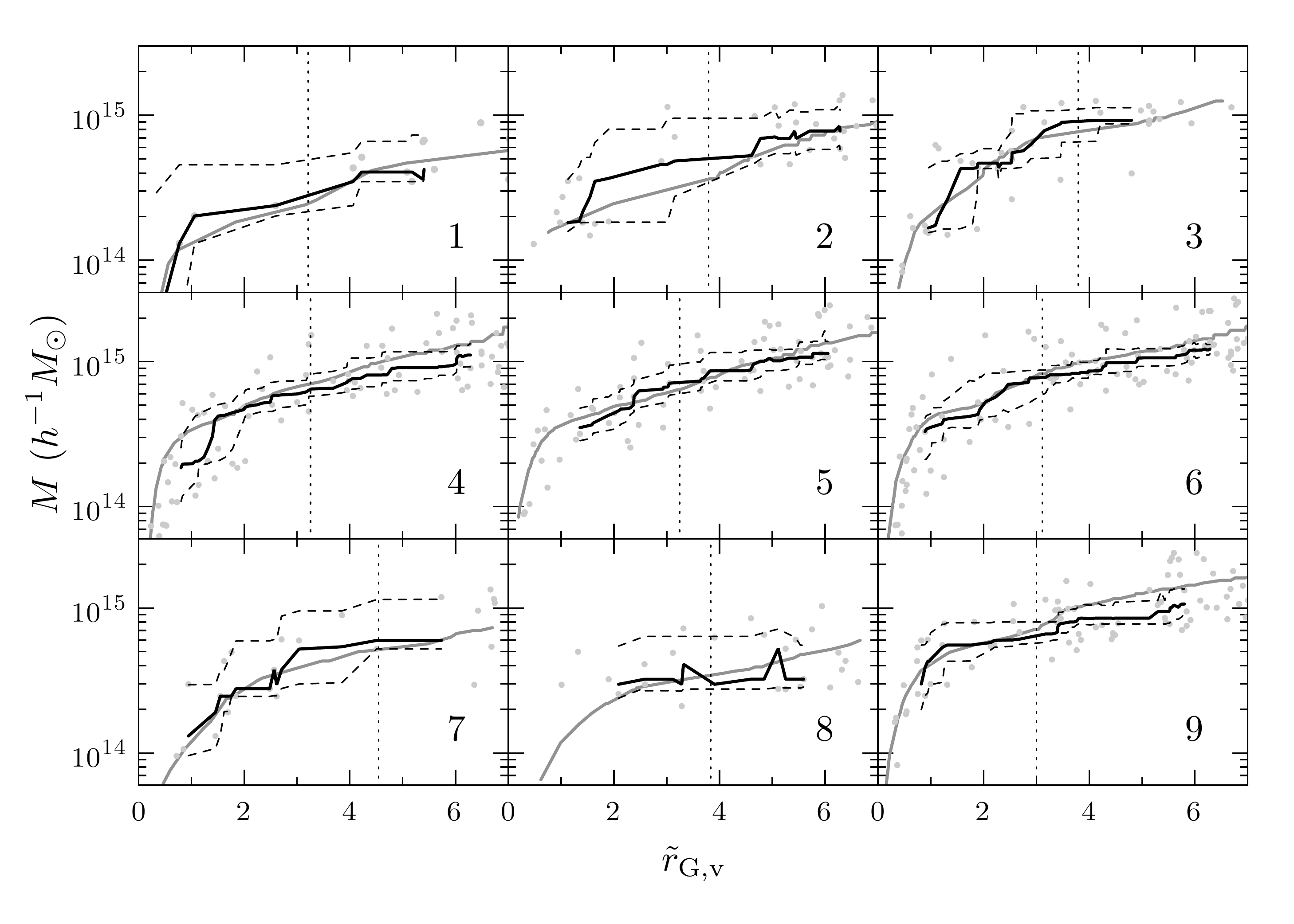}
\caption[]{Mass profiles for nine simulated clusters from our data catalogue. In each panel, we represented the actual cluster mass profile $M_g$ (grey solid lines), the distribution of $M_{\rmn{raw};g}$ values (grey dots) and the extimated mass profile $M_{\rmn{corr};g}$ (black solid lines) together with its uncertainty band (black dashed lines; see Section \ref{sec:mass_est}). The turnaround radii $r_\rmn t$ are also shown (vertical dotted lines).}
\label{fig:SC9}
\end{figure*}

Fig.~\ref{fig:SC9} shows the mass profiles of the nine simulated clusters. The grey solid lines represent the actual profiles $M_g$, while the black solid lines and dashed lines represent the estimated profile $M_{\rmn{corr};g}$, together with its uncertainty band. We adopted the $2\sigma$ uncertainty on the running median as the uncertainty on this estimate, since it was observed to mimic well the dispersion in the actual distribution of $M_{\rmn{G};g}$ values. The distributions of values $M_{\rmn{raw};g}$ are also shown with grey dots. This subset of nine clusters gives a good representation of the common results obtained when estimating the mass of single clusters from the whole catalogue. In most cases, our technique is able to produce a mass profile extended from the virialization core up to the extreme outskirts of clusters (clusters $1$, $2$, $4$, $6$, $7$ and $9$), even when only a few FGs are identified (clusters $1$ and $7$). If the FGs are not evenly distributed, the estimated profile turns out to be radially limited (clusters $3$ and $8$) but the estimate is generally correct within the covered interval. The definition of the FR, which is the same for all clusters (as described in Section \ref{subsec:FR}) does not mask the individual features of different clusters, which are correctly reconstructed through the features of the FG subset. In fact, our technique is able to discern the presence of remarkable changes of slope in the mass profiles (clusters $1$, $2$, $3$, $6$ and $9$), which cannot be predicted using the results of CMM08. In some cases, the procedure of correction and smoothing produces an undesired plateau in the outermost regions (clusters $9$). Only in few cases, the poor quality of data prevents our technique from properly reconstructing the mass profile (cluster $8$).

\begin{table*}
\centering
\caption[]{Estimated virialization masses and turnaround masses for the nine simulated clusters in Fig.~\ref{fig:SC9}, compared with the corresponding values extracted from the simulation.
}
\begin{tabular}{ccccc}
\hline
cluster & $M_\rmn{v}\, (10^{14} M_\odot)$ & $M_\rmn{corr,v}\, (10^{14} M_\odot)$ & $M_\rmn{t}\, (10^{14} M_\odot)$ & $M_\rmn{corr,t}\, (10^{14} M_\odot)$\\
\hline
$1$ & $1.37$ & $1.87_{-0.85}^{+2.69}$ & $2.48$ & $2.86_{-0.67}^{+2.14}$\vspace{0.1cm}\\
$2$ & $1.75$ & $1.80_{-0.35}^{+1.33}$ & $3.87$ & $5.05_{-1.51}^{+4.52}$\vspace{0.1cm}\\
$3$ & $2.07$ & $1.70_{-0.13}^{+2.75}$ & $8.13$ & $9.24_{-0.49}^{+2.08}$\vspace{0.1cm}\\
$4$ & $3.36$ & $1.98_{-0.60}^{+1.89}$ & $7.04$ & $6.44_{-0.67}^{+1.74}$\vspace{0.1cm}\\
$5$ & $3.64$ & $3.22_{-0.59}^{+1.17}$ & $6.40$ & $7.19_{-1.17}^{+2.44}$\vspace{0.1cm}\\
$6$ & $4.08$ & $3.53_{-1.14}^{+0.89}$ & $8.25$ & $7.77_{-1.87}^{+0.98}$\vspace{0.1cm}\\
$7$ & $1.16$ & $1.38_{-0.41}^{+1.58}$ & $5.26$ & $5.97_{-0.75}^{+5.46}$\vspace{0.1cm}\\
$8$ & $1.17$ & $2.25_{-0.89}^{+0.96}$ & $3.45$ & $3.09_{-0.33}^{+3.28}$\vspace{0.1cm}\\
$9$ & $4.11$ & $4.43_{-1.47}^{+2.33}$ & $7.12$ & $6.45_{-0.71}^{+1.57}$\vspace{0.1cm}\\
\hline
\end{tabular}
\label{tab:SC9}
\end{table*}

The values of  the virialization mass and the turnaround mass estimated for the nine simulated clusters in Fig.~\ref{fig:SC9} are listed in Table \ref{tab:SC9}, along with the corresponding values extracted from the simulations. In most cases, the agreement is fairly good, even when the overall profile is poorly reconstructed, thanks to the linear fitting algorithm used to compute $M_\rmn{corr,v}$ and $M_\rmn{corr,t}$.

\begin{figure*}
\centering
\includegraphics[width=16cm]{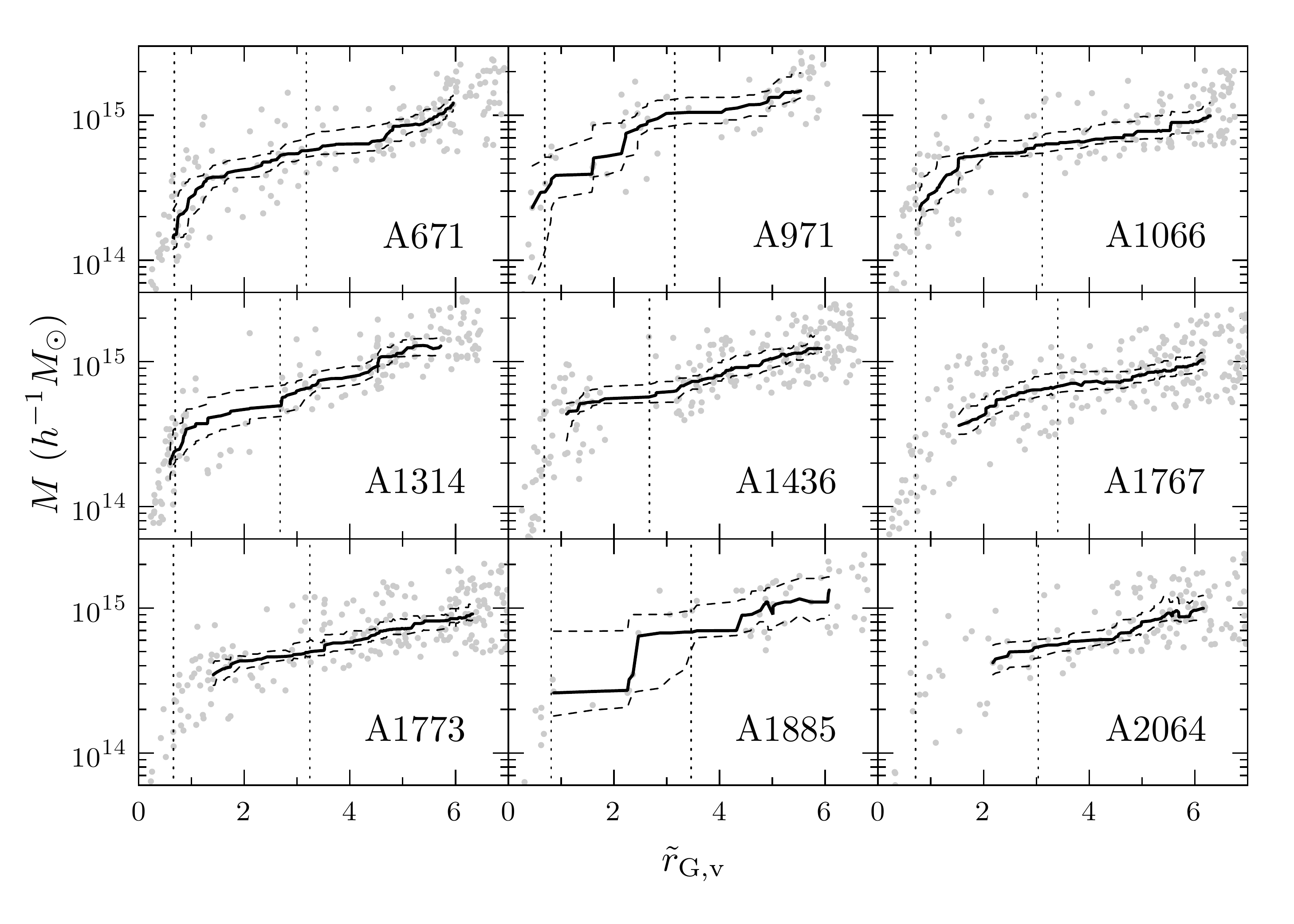}
\caption[]{Mass profiles for nine observed clusters from the CIRS Catalogue (RD). In each panel, we represented the distribution of $M_{\rmn{raw};g}$ values (grey dots) and the extimated mass profile $M_{\rmn{corr};g}$ (black solid lines) together with its uncertainty band (black dashed lines, see section \ref{sec:mass_est}). The virialization radii $r_{200}$ and the turnaround radii $r_\rmn t$ are also shown (vertical dotted lines, leftmost and rightmost in each panel, respectively).
}
\label{fig:SC9_CIRS}
\end{figure*}

\begin{table*}
\centering
\caption[]{Estimated virialization masses and turnaround masses for the nine CIRS clusters in Fig.~\ref{fig:SC9_CIRS}, compared with the corresponding values obtained by RD.}
\begin{tabular}{lcccc}
\hline
cluster & $M_\rmn{200}\, (10^{14} M_\odot)$ & $M_\rmn{corr,200}\, (10^{14} M_\odot)$ & $M_\rmn{t}\, (10^{14} M_\odot)$ & $M_\rmn{corr,t}\, (10^{14} M_\odot)$\\
\hline
A671 & $3.23\pm 1.02$ & $1.50_{-0.29}^{+0.82}$ & $4.40\pm 1.50$ & $5.70_{-0.56}^{+1.52}$\vspace{0.1cm}\\
A971 & $4.46\pm 0.89$ & $2.96_{-1.81}^{+2.15}$ & $4.95\pm 1.05$ & $10.4_{-1.86}^{+2.56}$\vspace{0.1cm}\\
A1066 & $4.68\pm 0.51$ & $1.98_{-0.91}^{+0.49}$ & $7.93\pm 1.06$ & $6.25_{-0.78}^{+1.02}$\vspace{0.1cm}\\
A1314 & $1.72\pm 0.44$ & $2.42_{-0.39}^{+1.04}$ & $2.28\pm 0.69$ & $4.96_{-0.78}^{+1.88}$\vspace{0.1cm}\\
A1436 & $0.86\pm 0.22$ & $3.27_{-2.91}^{+1.45}$ & $1.83\pm 0.67$ & $5.72_{-0.51}^{+1.24}$\vspace{0.1cm}\\
A1767 & $6.03\pm 2.36$ & $2.46_{-0.61}^{+1.33}$ & $11.3\pm 4.88$ & $6.76_{-0.65}^{+1.56}$\vspace{0.1cm}\\
A1773 & $1.98\pm 1.90$ & $1.72_{-0.88}^{+0.73}$ & $3.30\pm 3.18$ & $4.98_{-0.34}^{+1.01}$\vspace{0.1cm}\\
A1885 & $4.50\pm 1.24$ & $2.59_{-0.81}^{+4.33}$ & $4.46\pm 1.23$ & $6.82_{-1.76}^{+2.67}$\vspace{0.1cm}\\
A2064 & $1.65\pm 0.63$ & $0.23_{-0.23}^{+2.26}$ & $4.27\pm 1.48$ & $5.39_{-0.87}^{+0.68}$\vspace{0.1cm}\\
\hline
\end{tabular}
\label{tab:SC9_CIRS}
\end{table*}

Fig.~\ref{fig:SC9} shows the mass profiles of the nine CIRS clusters. These clusters were chosen among the ones with the highest radius of complete radial coverage in the survey. As before, the black solid lines and dashed lines represent the estimated profile $M_{\rmn{corr};g}$, together with its uncertainty band, while the distributions of values $M_{\rmn{raw};g}$ are shown with grey dots. The virialization radii $r_\rmn v$ of these clusters, needed to normalized the galaxy radial distribution, were computed iteratively as the radii where the overdensity estimated by our technique equals the virialization value $\delta_v\simeq 101/\Omega_0-1$ (\citealt{BN}; CMM08)\footnote{An alternative approach for computing $r_\rmn v$ is provided by the SCM. According to the results CMM08 and CMM10, the ratio between the turnaround radius and the virialization radius is approximately constant: $r_\rmn t/r_\rmn v\simeq 3.5$. Therefore, $r_\rmn v$ can be obtained from $r_\rmn t$, defined as the radius were the estimated overdensity equals the turnaround value $\delta_\rmn t\simeq 15$. This approach is suitable for estimating the values of $r_\rmn v$ and $M_\rmn v$ in poorly populated clusters, when no FGs are identified within the virialization core.}.The CIRS clusters are generally more populated than those extracted from our simulation, allowing a lower uncertainty in the mass profile estimation. A fairly good agreement with the estimates of RD is generally observed at all radii. Compared to the results of RD, our estimate is generally less accurate in the cluster core and better accurate in the outskirts. In most cases, our technique is able to reconstruct the mass profiles well beyond the turnaround radius.

The values of virialization mass and turnaround mass estimated for the nine CIRS clusters in Fig.~\ref{fig:SC9_CIRS} are listed in Table \ref{tab:SC9_CIRS}, along with the corresponding values estimated by RD. Only in this case, we adopted as virialization mass the value $M_{\rmn{corr},200}$, which corresponds to the commonly used overdensity value $\delta_{200}=200/\Omega_0-1$, to allow better comparison with the values $M_{200}$ estimated by RD. We point out that such estimation of $M_{\rmn{corr},200}$ is meant only to provide a reference value and must not be regarded as reliable, because $r_{200}$ is smaller $r_\rmn v$ and falls outside the effectiveness interval of our tecnique. Nevertheless, the values of $M_{200}$ and $M_{\rmn{corr},200}$ are in quite good agreement within the uncertainties. A fairly good agreement is observed also between $M_\rmn t$ and $M_\rmn{corr,t}$. In some cases (clusters A971, A1436, A1885) our estimates of the turnaround mass are significantly higher than those of RD. This is not unexpected, since the profile of these clusters as estimated by RD are truncated before actually reaching $r_\rmn t$, resulting in a possible systematic underestimation of $M_\rmn t$. 
\section{Discussion and conclusion}\label{sec:concl}

We discussed a new approach to estimate mass profiles and masses of galaxy clusters, based on the relation existing between the cluster overdensity and the infall velocity of member galaxies \citep{RG}. Our technique is simple, it only needs to know sky-plane positions and the line-of-sight velocities of the galaxies, and it can be applied to observed clusters.

Our analysis was performed and checked on a simulated catalogue of clusters \citep{Borgal, Bivial} in the theoretical framework of the SCM which is widely accepted in literature (CMM10). We found the existence of an FR in the redshift space and we demonstrated that the galaxies belonging to this region, i.e.~the FGs, are able to effectively identify the mass profiles of clusters and to measure the corresponding total masses.

Our technique consists in:
\begin{enumerate}
\p Identification of the FG subset via the identification of the FR in the redshift space;
\p Detection of the mass profiles using the redshift-space coordinates of FGs.
\end{enumerate}
We showed that it is possible to estimate the cluster mass profiles from $1$ up to $7$ virialization radii, within a tipical uncertainty factor of 1.5, for more than 90 per cent of clusters. The technique is reliable even with few identified Fair galaxies. 

Our technique was tested to be accurate for clusters in the mass range provided by our simulation, corresponding to virialization masses between $8.0\times10^{13}\,h^{-1}M_\odot$ and $1.3\times 10^{15}\,h^{-1}M_\odot$. In the present paper, we chose for the cosmological matter density parameter the value $\Omega_{\rmn M,0}=0.3$, which is also the value adopted by the simulation we used. Our results can be adapted to different values $\Omega_{\rmn M,0}$, ranging from $0.2$ to $0.4$. 

We applied our technique in a subset of clusters taken from the CIRS Catalogue (RD). The mass profiles and the masses we obtained are in fairly good agreement with previous literature values.

In the future, we aim to apply the present technique to clusters of large observational catalogues.

\section*{Acknowledgments}

We wish to thank Stefano Borgani for making available to us the simulated data, Andrea Biviano and Marisa Girardi for providing the simulated galaxy catalogue, and all of them for the useful discussions and insightful advices. We are also particularly thankful to Ken Rines and Antonaldo Diaferio for making available to us the CIRS catalogue in electronic form. We wish to thank the anonymous referee for the discussion and for the useful suggestions.

\bsp

\end{document}